\documentstyle[epsf,rotate,12pt]{article}
\newlength{\dinwidth}
\newlength{\dinmargin}
\newcommand{\ra}{\rightarrow}
\newcommand{\Li}{\mbox{Li}}
\newcommand{\BR}{\mbox{BR}}
\newcommand{\BGAMAXS}{B \ra X _{s} + \gamma}

\newcommand{\BBGAMAXS}{\BR (B \ra  X _{s} + \gamma)}

\newcommand{\GGAMAXS}{\Gamma (B \ra  X _{s} + \gamma)}

\newcommand{\BGAMAS}{b \ra s + \gamma}

\newcommand{\BGAMAGS}{ b \ra s  + g + \gamma}

\newcommand{\bsgam}{\ $b \to s+ \gamma$}

\newcommand{\bsggam}{\ $b \to s+ \gamma+ g$}

\newcommand{\absvcb}{\vert V_{cb}\vert}

\newcommand{\absvts}{\vert V_{ts}\vert}

\newcommand{\ba}{\begin{array}}
\newcommand{\ea}{\end{array}}
\newcommand{\be}{\begin{equation}}
\newcommand{\ee}{\end{equation}}
\newcommand{\bea}{\begin{eqnarray}}
\newcommand{\eea}{\end{eqnarray}}

\def\bra{\langle}
\def\ket{\rangle}

\def\a{\alpha}
\def\b{\beta}
\def\g{\gamma}

\def\p{\pi}

\def\ep{\varepsilon}

\def\l{\lambda}
\def\m{\mu}
\def\n{\nu}
\def\G{\Gamma}

\def\to{\rightarrow}

\begin{document}
\thispagestyle{empty}
\addtocounter{page}{-1}
\begin{flushright}
DESY 95--117\\
SLAC-PUB-95-6940\\
hep-ph/9506374\\
June 1995
\end{flushright}
\vspace*{3cm}
\centerline{\Large\bf Photon Energy Spectrum in $B \to X_s +\gamma$}
\vspace*{0.7cm}
\centerline{\Large\bf and   comparison with Data}
\vspace*{2.0cm}
\centerline{\large\bf A. Ali}
\vspace*{0.5cm}
\centerline{\large\bf Deutsches Elektronen Synchrotron DESY, Hamburg,
                                                      Germany}
\vspace*{0.5cm}
\centerline{\large\bf C. Greub \footnote{
                   Supported by Schweizerischer Nationalfonds.}}
\vspace*{0.5cm}
\centerline{\large\bf SLAC Theory Division, Stanford University, USA}
\vspace*{3cm}
\centerline{\Large\bf Abstract}
\vspace*{1cm}
     A comparison of the inclusive photon energy spectrum
in the radiative decay $\BGAMAXS$, measured recently by the CLEO collaboration,
with the standard model is presented, using a $B$-meson wave function model
and improving earlier perturbative QCD-based
computations of the same.
The dependence of the photon energy
spectrum on the non-perturbative
model parameters,
$p_F$, the $b$-quark Fermi momentum in the $B$ hadron,
 and $m_q$, the spectator quark mass,
is explicitly shown, allowing a comparison of these parameters with
the ones obtained from the analysis
of the lepton energy spectrum in semileptonic $B$ decays. Taking into
account present uncertainties, we estimate $\BBGAMAXS = (2.55 \pm 1.28)
\times 10^{-4}$ in the standard model,
 assuming $\absvts/\absvcb= 1.0$. Comparing this
with the CLEO measurement $\BBGAMAXS = (2.32 \pm 0.67) \times 10^{-4}$
implies $\absvts/\absvcb= 1.1 \pm 0.43$,
in agreement with the CKM unitarity.
\vskip1cm
\newpage
\section{Introduction}
\label{sec:introd}

    Recently, the CLEO collaboration has reported the first measurement of
the photon energy
spectrum in the decay $\BGAMAXS$ \cite{CLEOrare2}, following the measurement
of the exclusive decay mode $B \to K^* + \gamma$ reported in 1993 by the same
collaboration \cite{CLEOrare1}.
The inclusive branching ratio and the photon energy spectrum allow a less
model-dependent comparison with the underlying theory, more specifically the
 standard model (SM), as compared to the exclusive decay modes which require
additionally decay form factors.
 The CLEO data have been compared in \cite{CLEOrare2}
with
the SM-based theoretical computations presented in \cite{ag1} - \cite{ag3},
allowing to draw the conclusion that
 agreement between theoretical predictions and experiment is
good, given that large uncertainties exist in both.
In particular, the measured branching ratio $\BBGAMAXS = (2.32 \pm 0.67)
\times 10^{-4}$ \cite{CLEOrare2} is in agreement with the
SM-based estimates in
\cite{ag1,ag3}, as well as with the ones in \cite{Buras94,Ciuchini94}.

     In this letter, we would like to report on an improved calculation
of the photon energy spectrum compared to what we have presented earlier
and which has been used in the CLEO analysis \cite{CLEOrare2}.
The main theoretical difference lies in the inclusion of the complete
operator basis $O_1,...,O_8$ for the effective Hamiltonian,
 defined below, in the computation of the partonic
processes $\BGAMAS$ and $\BGAMAGS$, and in using the complete
 leading-logarithmic
computations of the anomalous dimension matrix presented in \cite{Ciuchini}.
In contrast, our previous calculations were done in the truncated
approximation,
 where we had dropped the effects of the four-Fermi operators
$O_3,...,O_6$ and the chromomagnetic
 operator $O_8$ in the computation of the contributions
from $\BGAMAGS$. In addition, use was made of the anomalous dimension
matrix derived in \cite{Grinstein90}. The other ingredient of our
calculation, namely a specific
 $B$-meson wave function model
\cite{Alipiet,ACCMM} to incorporate the non-perturbative effects on spectra,
 remains unaltered. However, since data are now
 available,
we fit the normalized photon energy spectrum with the improved theoretical
framework to determine from the shape the non-perturbative parameters of
the wave function model being used, namely the Fermi motion parameter $p_F$
and the spectator quark mass $m_q$. These in turn determine within the model
the $b$-quark
mass. There is considerable theoretical interest in these
parameters, in particular $p_F$,
which is a good measure of the kinetic energy of the $b$ quark in the
$B$ meson, and the $b$-quark mass.
 Since a similar framework has also been used, in conjunction with the
perturbative QCD-improved parton model, in the
analysis of the lepton energy spectrum in inclusive $B$-meson decays
\cite{CLEOrare2}, it
is  interesting to compare the model parameters obtained from the
two measurements. While, admittedly, present errors are large preventing
us from drawing sharp conclusions, some valuable
insight on the shape parameters and normalization can
already  be obtained and we
quantify this information.

     This letter is organized as follows: In section 2, we briefly
summarize the effective Hamiltonian for the decay $\BGAMAXS$ and present
the Wilson coefficients numerically. Section 3 contains an anatomy of the
partonic processes $\BGAMAS$ and $\BGAMAGS$, where the essential steps
in the derivation of the matrix elements are given. The photon energy
spectrum at the partonic level is derived in section 4, including a
discussion of the Sudakov behaviour in the end-point region. Section 5
summarizes the $B$-meson wave function model \cite{Alipiet,ACCMM}.
Numerical results for
the branching ratio $\BBGAMAXS$ in the SM, the ratio
of the Cabibbo-Kobayashi-Maskawa (CKM) matrix elements $\absvts/\absvcb$,
and fits of the CLEO photon energy spectrum from $\BGAMAXS$ with our
theoretical estimates, yielding a $\pm 1 \sigma$-contour in the parameter
space $(p_F,m_q)$ are given in section 6.


\section{Effective Hamiltonian for the decay $\BGAMAXS$}
\label{sec:effham}
The framework  we use here is that of an effective theory with
five quarks, obtained by integrating out the
heavier degrees of freedom,
which in the standard model are the top quark and the $W$-boson.
A complete set of dimension-6 operators relevant for the processes
\bsgam  ~and \bsggam ~is contained in the effective Hamiltonian
\begin{equation}
\label{heff}
H_{eff}(b \to s \gamma)
       = - \frac{4 G_{F}}{\sqrt{2}} \, \lambda_{t} \, \sum_{j=1}^{8}
C_{j}(\mu) \, O_j(\mu) \quad ,
\end{equation}
where
$G_F$ is the Fermi constant
coupling constant and
$C_{j}(\mu) $ are the Wilson coefficients evaluated at the scale $\mu$,
and $\lambda_t=V_{tb}V_{ts}^*$ with $V_{ij}$ being the
CKM matrix elements.
The operators $O_j$ read
\bea
\label{operators}
O_1 &=& \left( \bar{c}_{L \b} \g^\m b_{L \a} \right) \,
        \left( \bar{s}_{L \a} \g_\m c_{L \b} \right)\,, \nonumber \\
O_2 &=& \left( \bar{c}_{L \a} \g^\m b_{L \a} \right) \,
        \left( \bar{s}_{L \b} \g_\m c_{L \b} \right) \,,\nonumber \\
O_3 &=& \left( \bar{s}_{L \a} \g^\m b_{L \a} \right) \, \left[
        \left( \bar{u}_{L \b} \g_\m u_{L \b} \right) + ... +
        \left( \bar{b}_{L \b} \g_\m b_{L \b} \right) \right] \,,
        \nonumber \\
O_4 &=& \left( \bar{s}_{L \a} \g^\m b_{L \b} \right) \, \left[
        \left( \bar{u}_{L \b} \g_\m u_{L \a} \right) + ... +
        \left( \bar{b}_{L \b} \g_\m b_{L \a} \right) \right] \,,
        \nonumber \\
O_5 &=& \left( \bar{s}_{L \a} \g^\m b_{L \a} \right) \, \left[
        \left( \bar{u}_{R \b} \g_\m u_{R \b} \right) + ... +
        \left( \bar{b}_{R \b} \g_\m b_{R \b} \right) \right] \,,
        \nonumber \\
O_6 &=& \left( \bar{s}_{L \a} \g^\m b_{L \b} \right) \, \left[
        \left( \bar{u}_{R \b} \g_\m u_{R \a} \right) + ... +
        \left( \bar{b}_{R \b} \g_\m b_{R \a} \right) \right] \,,
        \nonumber \\
O_7 &=& (e/16\p^{2}) \, \bar{s}_{\a} \, \sigma^{\m \n}
      \, (m_{b}(\mu)  R + m_{s}(\mu)  L) \, b_{\a} \ F_{\m \n} \,,
        \nonumber \\
O_8 &=& (g_s/16\p^{2}) \, \bar{s}_{\a} \, \sigma^{\m \n}
      \, (m_{b}(\mu)  R + m_{s}(\mu)  L) \, (\l^A_{\a \b}/2) \,b_{\b}
      \ G^A_{\m \n} \quad ,
        \nonumber \\
\eea
where $e$ and $g_s$ are the electromagnetic and the strong
coupling constants, respectively. In the magnetic moment type
operators $O_7$ and $O_8$, $F_{\m \n}$ and $G^A_{\m \n}$
denote the electromagnetic and the gluonic field strength
tensors, respectively.
$L=(1-\g_5)/2$ and $R=(1+\g_5)/2$
denote the left and right-handed projection operators.
We note here that the explicit mass factors in $O_7$
and $O_8$ are the running quark masses.
QCD corrections to the decay rate for $b \to s \g$
bring in
large logarithms of the form $\a_s^n(m_W) \, \log^m(m_b/M)$,
where $M=m_t$ or $m_W$ and $m \le n$ (with $n=0,1,2,...$).
To get a reasonable result, one
should resum at least the leading logarithmic contribution
(i.e. $m=n$) to all orders.
Using the renormalization group equation the Wilson coefficient
can be calculated at the scale $\mu \approx m_b$ which
is the relevant scale for $B$ decays.
At this scale the large logarithms are contained in the
Wilson coefficients.
As we are working in this paper to leading logarithmic precision,
it is sufficient to know the leading order
anomalous dimension matrix and the matching
$C_i(\m=m_W)$ to lowest order (i.e., without
QCD corrections) \cite{InamiLim}.
The $8 \times 8$ matrix is given in
\footnote{The results given here for the entries concerning
$O_7$ and $O_8$ correspond to the naive dimensional regularization
scheme (NDR), which we use in the calculation
of all the matrix elements.}
\cite{Ciuchini} and the
  Wilson coefficients
are explicitly listed in
\cite{Buras94,AGM94}.
The numerical values of the Wilson
coefficients that we use in our calculations are given in
 table 1.
For subsequent discussion we define two
effective Wilson coefficients $C_7^{eff}(\mu)$ and $C_8^{eff}(\mu)$
below and give their numerical values in table 1:
\begin{eqnarray}
\label{C78eff}
C_7^{eff} &\equiv & C_7 + Q_d \, C_5 + 3 Q_d \, C_6 \quad , \nonumber\\
C_8^{eff} &\equiv & C_8 + C_5 \quad .
\end{eqnarray}
\begin{table}[htb]
\label{wcmudep}
\begin{center}
\begin{tabular}{| c | c | r | r | r | }
\hline
 $C_i(\mu)$ & $\mu=m_W$ & $\mu=10.0$ GeV
                              & $\mu=5.0$ GeV
                              & $\mu=2.5$ GeV\\
\hline \hline
$C_1$ & $0.0$ & $-0.158$ & $-0.235$ & $-0.338$ \\
$C_2$ & $1.0$ & $1.063$ & $1.100$ & $1.156$ \\
$C_3$ & $0.0$ & $0.007$ & $0.011$ & $0.016$ \\
$C_4$ & $0.0$ & $-0.017$ & $-0.024$ & $-0.034$ \\
$C_5$ & $0.0$ & $0.005$ & $0.007$ & $0.009$ \\
$C_6$ & $0.0$ & $-0.019$ & $-0.029$ & $-0.044$ \\
$C_7$ & $-0.193$ & $-0.290$ & $-0.333$ & $-0.388$ \\
$C_8$ & $-0.096$ & $-0.138$ & $-0.153$ & $-0.171$ \\
$C_7^{eff}$ & $-0.193$ & $-0.273$ & $-0.306$ & $-0.347$ \\
$C_8^{eff}$ & $-0.096$ & $-0.132$ & $-0.146$ & $-0.162$ \\
\hline
\end{tabular}
\end{center}
\caption{Wilson coefficients $C_i(\mu )$
                                   at the scale $\mu=m_W=80.33$ GeV
(``matching conditions") and at three other scales, $\mu = 10.0$ GeV,
$\mu =5.0$ GeV and $\mu = 2.5$ GeV,
 evaluated with two-loop $\beta$-function and the
 leading-order anomalous-dimension matrix. The entries correspond to the top
quark mass
 $\overline{m_t}(m_t^{pole})=170$ GeV (equivalently,
$m_t^{pole}= 180 $ GeV) and the QCD parameter with 5 flavours
 $\Lambda_{5} = 195 $ MeV
(equivalently, $\alpha_s(m_Z^2)=0.117$), both in the $\overline{MS}$
scheme.}	 \label{tab3}
\end{table}
\section{An anatomy of the processes $b \to s \g$ and $b \to s \g g$}
\label{sec:bsgamma}
We summarize the principal points of the derivation of the photon
energy  spectrum in the decay $b \to s \gamma (+ g)$ in this letter and
refer to \cite{AGlong} for technical details.
As it is the process $b \to s \g g $ which leads to a nontrivial
photon energy spectrum at the partonic level,
we strive at taking  it fully into account.
The amplitude for this decay suffers from infrared singularities as
the gluon or the photon
          energy goes to zero. The topology of the decay $ b \to s \g g$
resembles that of the two-body decays $b \to s \g$ and $b \to s g$
in the limit as $E_g \to 0$ or $E_{\g} \to 0$, respectively.
The singular configurations in $b \to s \g g$ are cancelled
in a distribution sense in the photon energy spectrum if one
also takes into account virtual corrections to the two-body process
$b \to s g$ and $b \to s \g$, order by order in perturbation theory.
We will take into account only those virtual correction diagrams
which are needed  to cancel the infrared singularities
from $b \to s \g g$.

We first discuss the result for $b \to s \g$. It turns out that
the effects of the four-Fermi operators to $b \to s \g$ can
be absorbed into a redefinition of the coefficient $C_7 \to C_7^{eff}$,
with $C_7^{eff}$ defined in eq. (\ref{C78eff}).
This not only holds for the $b \to s \g$ amplitude
without virtual corrections
but also for those infrared-sensitive virtual
corrections which we have mentioned above.
One therefore has to calculate the matrix element
of $C_7^{eff} O_7$ for $b \to s \g$ including virtual corrections.
The result of this calculation,
which was derived in $d=4-2 \epsilon$ dimensions,
can be expressed as \cite{ag1}:
\be
\label{gamma7virt}
\Gamma_{7}^{virt} = \Gamma_{7,sing}^{virt} +
\Gamma_{7,fin}^{virt} ~,
\ee
where we have split this quantity into an
infrared-finite and an infrared-singular part:
\begin{equation}
\label{virtsing}
\Gamma_{7,sing}^{virt} = - \frac{m_b^5}{96 \pi^5} \,
(1-r)^{3}\, (1+r) \,
|C_7^{eff} \, G_F \, \lambda_{t}|^{2} \,
 \a_{em} \, \a_s \,
\frac{\left( \frac{4 \pi \mu^2}{m_b^2} \right)^{2 \epsilon}}{
\Gamma(2-2\epsilon)}   \,
 \ \left[ \frac{4}{\epsilon} +
\frac{2(1+r)}{1-r} \, \frac{\log r}{\epsilon} \right] \,,
\end{equation}
and
\begin{eqnarray}
\label{virtfin}
\Gamma_{7,fin}^{virt} &=& \frac{m_b^5}{32 \pi^4} \,
(1-r)^{3}\, (1+r) \,
|C_7^{eff} \, G_F \, \lambda_{t}|^{2} \,
 \a_{em} \,
(1 + \frac{\a_s}{3 \pi} \, \tau)   \,, \\
\label{tau}
\tau &=&  \left\{
\left( \frac{1+r}{1-r} \right) \left( \log^{2} r - 4 \log r
+ 4 \log r \, \log(1-r)
\right) - 8 + 3 \log r
+ 8 \log (1-r)  - 2 \log \frac{\mu^2}{m_b^2}
\right\}  ~.\nonumber \\
\end{eqnarray}
The quantity $r$ is defined as $r=(m_s/m_b)^2$.
The infrared singularity (\ref{virtsing})
will be cancelled when taking into account
the gluon bremsstrahlung diagrams involving the operator $O_7$.

As we already noted, the process $b \to s \g g$ has also an infrared
singularity as the photon becomes soft. These
singularities are cancelled analogously by virtual
photon corrections to the matrix element for $b \to s g$, i.e. in
$C_8^{eff} O_8$.
The result is
\bea
\label{virt8}
\G_8^{virt} &=& \G_{8,sing}^{virt}
+ \G_{8,fin}^{virt}  \,, \\
\label{virtsing8}
        \Gamma_{8,sing}^{virt} &=& - \frac{m_b^5}{96 \pi^5} \,
(1-r)^{3}\, (1+r) \,
|Q_d \, C_8^{eff} \, G_F \, \lambda_{t}|^{2} \,
 \a_{em} \, \a_s \,
\frac{\left( \frac{4 \pi \mu^2}{m_b^2} \right)^{2 \epsilon}}{
\Gamma(2-2\epsilon)}   \,
\, \left[ \frac{4}{\epsilon} +
\frac{2(1+r)}{1-r} \, \frac{\log r}{\epsilon} \right]        \\
\label{virtfin8}
\Gamma_{8,fin}^{virt} &=& \frac{m_b^5}{96 \pi^5} \,
(1-r)^{3}\, (1+r) \,
|Q_d \, C_8^{eff} \, G_F \, \lambda_{t}|^{2} \,
 \a_{em} \,  \a_s \, \tau \,,
\eea
where $\tau$ is given in eq. (\ref{tau}) and $Q_d=-1/3$.
A remark concerning the quark masses is in order here. When
calculating the matrix elements we have used the on-shell
subtraction prescription for the quark masses. Due to the explicit
factors of the running quark masses in the operators $O_7$ and
$O_8$, the $m_b^5$ factor contained in $\G_7^{virt}$
and $\G_8^{virt}$ given above should be replaced by the following product
\be
m_b^5 \longrightarrow m_{b,pole}^3 m_b(\mu)^2 \quad ,
\ee
where $m_{b,pole}$ and $m_b(\mu)$ denote the pole mass and the
running mass of the $b$ quark, respectively.
In actual practice, we identify all the masses $m_b$ in the various
intermediate expressions with $m_{b,pole}$ and multiply at the end the
so-derived  decay width $\GGAMAXS$
with a correction factor R:
\be
\label{rfactor}
R=(m_b(\mu)/m_{b,pole})^2 \quad ,
\ee
as also advocated in \cite{Buras94}.

We now take up the
matrix elements for the process $b \to s \g g$.
As the explicit expressions are too long to be presented here,
we only point out the basic structure and give
the complete formulae elsewhere \cite{AGlong}.
We first concentrate on the  contributions of the four-Fermi
operators $O_1, ..., O_6$. It turns out that the diagrams
which do not involve both gauge particle radiation from an internal
quark are either zero or can be absorbed into a redefinition
of the coefficients $(C_7, C_8) \to (C_7^{eff}, C_8^{eff})$.
The remaining case, in which both gauge particles
are emitted from the internal fermion line, is discussed now.
There are two such diagrams associated with each
four-Fermi operator, whose sum is ultraviolet
(and infrared) finite. We denote these matrix elements
by $M_i$
\be
M_i = \frac{4 G_F}{\sqrt{2}} \bra s \g g |C_i O_i |  b \ket \quad,
(i=1,2,..., 6) \quad .
\ee
Analogously, the matrix elements of $C_7^{eff} O_7$ and
$C_8^{eff} O_8$ are denoted by $M_7$ and $M_8$, respectively. Adding all
these contributions, the complete
matrix element for $b \to s \g g$ is denoted by M
\be
M = \sum_i^8 M_i \quad .
\ee
When squaring $M$ and summing over the polarizations and spins
of the particles, it turns out that only $|M_7|^2_\Sigma$ and
$|M_8|^2_\Sigma$ are infrared-singular. The other squared
amplitudes  and all
the interference terms are infrared-safe.
We therefore make the decomposition
\be
\label{mdec}
|M|^2_\Sigma = F + |M_7|^2_\Sigma + |M_8|^2_\Sigma \quad ,
\ee
where $F$ denotes the infrared-safe contributions.
%
\section{The photon energy spectrum at the partonic level}
In the following, it is useful to
define the dimensionless photon energy $x$ through the relation
\[ E_\g = \frac{m_b^2-m_s^2}{2 \, m_b} \, x \quad ; \]
$x$ then varies in the interval $[0,1]$.
The contribution of the $F$-term in eq. (\ref{mdec})
to the spectrum
$d\G_F/dx$ is completely finite and is worked out numerically.
As the contribution $d\G_7^{brems}/dx$ associated with
$|M_7|^2_\Sigma$ is  singular, we work it
out analytically in $d=4-2\ep$ dimensional phase space.
Making use of the abbreviations
$r=(m_s/m_b)^2$ and
$\xi=(1-r)x$ the quantity
$d\G_7^{brems}/dx$ can be written as
\bea
\label{gamma7}
\frac{d \G_7^{brems}}{dx} &=& C_\ep
(1-r)^{-4 \ep} \frac{x^{-2 \ep}}{(1-x)^{1+2\ep}} \, I_\ep(x) +
C_{\ep=0}
\left\{ \frac{x(2 x^2 - 5x -1)(1-r)}{1-\xi}   \right.
\nonumber \\
&& \left. + \frac{(1-r)x(1-x)(2x-1)}{(1-\xi)^2}
-2(1+x) \log(1-\xi) - I_{\ep=0}(x) \right\}
\quad ,
\eea
with
\bea
\label{explan}
C_\ep &=& m_b^5 \,
\frac{\left( \frac{4 \pi \mu^2}{m_b^2} \right)^{2 \ep}}{\G(2-2\ep)}
 \frac{|G_F \l_t C_7^{eff}|^2}{96 \pi^5} \,
\a_{em} \a_s \,  (1+r) (1-r)^3 \,,
\nonumber \\
I_\ep(x) &=& I_a(x) + I_b(x) \ep \nonumber \,, \\
I_a &=& -4 - \frac{4r}{1-\xi} - \frac{4}{\xi} (1+r)\log(1-\xi) \,,
\nonumber \\
I_b &=& \frac{8(1-\xi)}{\xi} \log(1-\xi) - \frac{4r}{1-\xi}
\log(1-\xi) -
\frac{8r}{1-\xi} + 8(1+r) \frac{\Li(\xi)}{\xi}
- \frac{2(1+r)}{\xi} \log^2(1-\xi)  ~.\nonumber \\
\eea
Here the symbol Li stands for the Spence function.
 The photon energy
spectrum away from the endpoint $x=1$ can
be obtained by taking the limit $\ep \to 0$ in
eqs. (\ref{gamma7}) and (\ref{explan}).
However, as the experimentally interesting
region is just near this endpoint one has to add
the virtual QCD corrections to the two-body
process $b \to s \g$, as already mentioned.
Before we discuss this further, we give the result
for $d\G_8^{brems}/dx$.

As $d\G_8^{brems}/dx$ has a non-integrable singularity
at  $x=0$ (i.e., $E_\g=0$), we should
work in $d$-dimensions as well. However,
as the photon energy spectrum is of no experimental interest at
very small energies, i.e. $E_\gamma \to 0$,
we remove the infrared regularization immediately. As for the branching
 ratio, we note that the dimensionally regularized total
integral $\G_8^{brems}$ can be obtained from the analogous expression for
$\G_7^{brems}$ by doing
obvious replacements.
 The result for
$d\G_8^{brems}/dx$ reads
\bea
\label{gamma8}
\frac{d \G_8^{brems}}{dx} &=& m_b^5 \,
 \frac{|G_F \l_t Q_d C_8^{eff}|^2}{96 \pi^5} \,
\a_{em} \a_s \, (1+r) (1-r)^3 \, (1-x)  \nonumber \\
&& \times \left\{ \frac{2[(1-r)x(x-2)+2(1+r)]}{x(1-x)(1-r)} \,
\log \frac{1-\xi}{r} + \right. \nonumber \\
&& \left. \frac{(1-r)(1-x)(1-2x)}{(1-\xi)^2}
- \frac{(1-r)(2x^2-x+1)}{1-\xi}  - \frac{8}{x} \right\} \quad .
\eea
Returning to the discussion of the infrared singularity
of the quantity $d\G_7^{brems}/dx$ in eq. (\ref{gamma7}), which occurs
for $x \to 1$, i.e., in the
experimentally interesting
region, we recall that
this singularity is cancelled in a distribution sense if one takes
into account the virtual QCD corrections to the tree level matrix
element of $O_7$ for the
two-body
process $b \to s \g$, given in eqs. (\ref{virtsing})
and (\ref{virtfin}).
Technically, it is useful to define the integrated quantity
\be
\label{gamma7s0}
\G_7(s_0) =
\int_{s_0}^{1} \, \left[ \, \frac{d\G_7^{brems}}{dx} +
               \, \G_7^{virt} \, \delta(1-x) \right] \, dx
\quad ,
\ee
in which the singularities cancel manifestly. The expressions
for $d\G_7^{brems}/dx$ and $\G_7^{virt}$ are given in eqs.
(\ref{gamma7}) and
(\ref{gamma7virt}), respectively.
The explicit expression for $\G_7(s_0)$ reads
\be
\label{eqgamma7}
\Gamma_7(s_0) = V \, ( 1 + \frac{\a_s}{3 \pi} \, \Omega) \,
\Theta(1 - s_0) \quad ,
\ee
where $\Theta$ is the step function and  $V$ is
\be
\label{prefactor}
V =  m_b^5 \, \frac{|G_F \l_t C_7^{eff}|^2}{32 \pi^4} \,
\a_{em} \, (1+r) (1-r)^3 \quad .
\ee
The expressions for $\Omega$  is identical with the
one that we had derived and presented earlier \cite{ag2}. The
difference lies in the normalization factor $V$ given above.
                                         We have now included the
complete operator basis, $O_1,...,O_8$, as opposed to our earlier
calculations \cite{ag1,ag2} where use was made of a truncated
basis neglecting the four-Fermi operators $O_3,...,O_6$. The difference
reflects itself in the coefficient $C_7$ used in \cite{ag1,ag2}, which
is now replaced by $C_7^{eff}$. The effective coefficient $C_7^{eff}$
calculated with the anomalous dimension matrix given above is
typically $10 \%$ smaller than $C_7$ used by us earlier. From
the definition of the quantity $\G_7(s_0)$ it is clear that
the photon energy spectrum is reconstructed by differentiation, i.e.,
\be
\label{reconstruct}
\frac{d\G_7(x)}{dx}(x) = - \frac{d\G_7(s_0)}{ds_0}|_{s_0 =x} \quad .
\ee
The end-point spectra, however,
show sensitivity to the leftover effects of the
(cancelled) infrared singularity, with
the photon-energy distribution
rising
very steeply near the end-point, $x \simeq 1$.
Furthermore, there is still a $\delta$-function contribution sitting at
$x=1$, stemming from differentiation of the $\Theta$ function
in eq. (\ref{eqgamma7}).
To remedy this, one often resorts to (an all order)
resummation of the leading (infrared) logarithms.
This resummation is done at the level of the quantity
$\Gamma_7(s_0)$, i.e., before one derives the spectrum
by the differentiation just described.
Although we are using for $\Omega$ the expression for
a non-zero strange quark mass (i.e. $r \ne 0$) in the numerics
\cite{ag2},
it is nevertheless instructive to look at the limit $r \to 0$.
Splitting $\Omega$ into two parts $\Omega=\Omega_1 + \Omega_2$
and using the notation $s_1=(1-s_0)$ we get
\bea
\label{omega}
\Omega_1 &=& -2 \log^2 s_1 - 7 \log s_1  \,, \nonumber \\
\Omega_2 &=& - 2 \log \frac{\m^2}{m_b^2}
- \frac{4\pi^2}{3} - 5 + 10s_1 + s_1^2 - \frac{2}{3}
s_1^3 + s_1 (s_1-4) \log s_1 \, .
\eea
Before giving the exponentiated analogue of eq. (\ref{eqgamma7}),
we point out
that the lepton energy spectrum in the
inclusive semileptonic decays $b \to u (+g) \ell \nu_\ell$,
 in the limit
of neglecting the final-state quark mass,  has a
similar structure \cite{CCM,JK}. The leading term, i.e. the
Sudakov double-log \cite{Sudakov}
in eq. (\ref{omega}),
also enters in the lepton energy spectrum in the mentioned semileptonic
decays  with
the same coefficient. However,
as shown here and earlier \cite{ag1,ag2} (and also discussed in
\cite{Neubert94}), the coefficients of the non-leading logarithmic and
power terms in the inclusive decay $\BGAMAXS$ and the semileptonic decay
$B \to X_u \ell \nu_\ell$ decay are different \footnote{The discrepancy
in the analytic results for the non-leading terms in \cite{CCM} and
\cite{JK} for the semileptonic
decay seems to have been settled in favour of \cite{JK}.}. Therefore,
the photon
energy spectrum in $\BGAMAXS$
  at the parton level can not be gotten from the
lepton energy spectrum in the decays $B \to X_u \ell \nu_\ell$.
For a recent discussion of the Sudakov-improved photon energy spectrum in
$\BGAMAXS$, see \cite{shifmangamma}.
Finally,
the exponentiated analogue of $\G_7(s_0)$ in eq.
(\ref{eqgamma7}) reads
\begin{equation}
\label{eqexp}
\G_7^{exp}(s_0) = V \, \left(1 + \frac{\a_s}{3\pi} \, \Omega_2
\right) \,
\exp \left( \frac{\a_s}{3\pi} \Omega_1 \right) \, \Theta(1-s_0)
\qquad .
\end{equation}
The explicit expressions for $\Omega_1$ and $\Omega_2$ for
$r \ne 0$ are given in \cite{ag2}.
The expression for the photon energy spectrum
can again be obtained by differentiation:
\be
\label{gammaexp}
\frac{d\G_7^{exp}(x)}{dx}(x) = - \frac{d\G_7^{exp}(
s_0)}{ds_0}|_{s_0 =x} \quad .
\ee
As exponentiation is required only near the end-point
                                        $x \to 1$, we use the
exponentiated form in the region $x > x_{crit}$ only. For
numerical calculations we take
$x_{crit} = 0.85$, where the transition
from one form to the other works smoothly.

To summarize, the final answer for the photon energy spectrum
for the process $b \to s \g (+g)$, where the $b$ quark decays
at rest, is given by
\begin{equation}
\label{decomp}
\frac{d\Gamma}{dx} =
\frac{d\Gamma_{F}}{dx} +
\Theta(x_{crit} - x)
                    \, \frac{d\Gamma_{7}^{brems}}{dx}
+ \Theta(x - x_{crit})
                      \, \frac{d\Gamma_{7}^{exp}}{dx}
+
                      \, \frac{d\Gamma_{8}^{brems}}{dx}
 \quad ,
\end{equation}
where $d\G_F/dx$ is too long to be given here and is relegated to
\cite{AGlong}; $d\G_7^{brems}/dx$, $d\G_7^{exp}/dx$ and
$d\G_8^{brems}/dx$
are given in eqs.~(\ref{gamma7}),
(\ref{gammaexp}) and
(\ref{gamma8}),
respectively. A remark is in order here: As it stands, eq.
(\ref{decomp}) has still a singularity at the experimentally
uninteresting point  $E_\g=0$, which is cancelled in a distribution
sense when taking into account the virtual photon correction
(see eq. (\ref{virt8}))
of the operator $O_8$. In principle, one could do a resummation
of the soft photons, in analogy with the treatment of the
resummation of soft gluons just discussed above. As it turns out
numerically, the contribution of photon energies below
50 MeV to the branching ratio is less than $1\%$, we just make
a cut there and do not resort to any resummation in the $E_\gamma \to 0$
limit.
%
\section{$B$-meson Wave function Effects
in $\BGAMAXS$}
\label{wavefunction}
In order to implement the $B$-meson bound state effects on the photon energy
spectrum, we continue to
use the wave-function model \cite{Alipiet,ACCMM} that we have used in our
earlier work \cite{ag2,ag3}.
 We recall that this model has also been used in calculating
the lepton energy spectra in inclusive $D$- and $B$-semileptonic decays.
Assuming that the $B$-meson wave function effects are universal, the
parameters of this model can be fixed from the semileptonic analysis
and one can make a parameter-free comparison of the photon energy
spectrum in $\BGAMAXS$
 with data. This is what has been done in the experimental
analysis of the data \cite{CLEOrare2}. However, in general, the
non-perturbative effects are likely to depend on the final-state quark mass,
which,
for example, is the case in the decays $\BGAMAXS$ and $B \to X_c
\ell \nu_\ell$.
It is, therefore, an interesting
question to ask if the non-perturbative model
being used by us consistently describes the lepton
and photon energy spectra in the inclusive decays $B \to X \ell \nu_\ell$
and $B \to X_s \gamma$, respectively. To answer this question quantitatively,
one has to implement the perturbative QCD effects, as discussed in the
previous section, incorporate the model, and confront
the resulting framework with data to
determine the model parameters with a well-defined statistical significance.
 This is what we shall undertake
in this paper.
 We note that this point has also
been investigated recently with a somewhat different non-perturbative (model)
and perturbative(QCD) input in \cite{shifmangamma}. Due to detailed
differences
in the underlying theoretical frameworks and in the analysis of data, no
direct quantitative comparison with this work is attempted here.

\indent
In the present model, which we refer to as the Fermi motion model,
 the $B$-meson consists of a $b$-quark and a spectator $q$ and
 the four-momenta of the
constituents are required to add up to the
four-momentum of the $B$-meson.
In the rest frame of the $B$-meson the $b$-quark and the spectator
fly
back-to-back with three momenta $\vec{p}_b=-\vec{p}_q \equiv \vec{p}$.
Energy conservation then implies the equation
\[ m_B = \sqrt{m_b^2 + \vec{p}^2} + \sqrt{m_q^2 + \vec{p}^2} \quad, \]
which can only hold for all values of $|\vec{p}|$,
if at least one of the
masses becomes momentum dependent. In this model the spectator quark
$m_q$ is chosen to be a fixed,
i.e. momentum-independent, parameter and the $b$-quark mass
becomes momentum dependent. This momentum dependent $b$-quark mass
is denoted by $W$ in the following and is given by
\begin{equation}
\label{lett14}
W^2(p) = {M_B}^2 + {m_q}^2 -2M_B \sqrt{p^2 + {m_q}^2} \quad .
\end{equation}
The $b$-quark, whose decays determine the dynamics, is given
 a non-zero momentum having a Gaussian
distribution, with the width determined by the
parameter $p_F$:
\begin{equation}
\label{lett13}
 \phi(p)= \frac {4}{\sqrt{\pi}{p_F}^3} \exp (\frac {-p^2}{{p_F}^2})
\quad ; \quad p = |\vec{p}|
\end{equation}
We note that this wave function is normalized such that
\[ \int_0^\infty \, dp \, p^2 \, \phi(p) = 1 \quad . \]
 The photon energy spectrum from
 the decay of the $B$-meson at rest is then given by
\begin{equation}
\label{lett15}
 \frac{d\Gamma}{dE_\gamma}= \int_0^{p_{max}} \, dp \, p^2 \, \phi(p)
  \frac {d\Gamma_b}{dE_\gamma}(W,p,E_\g) \quad ,
\end{equation}
where $p_{max}$ is the maximally allowed value of $p$ and
$ \frac{d\Gamma_b}{dE_\g}$
 is the photon energy spectrum from the decay of the $b$-quark in
flight, having a mass $W(p)$ and momentum $p$. This can be obtained by
Lorentz boosting the $b$-quark decay spectrum at rest, which has been
obtained in eq. (\ref{decomp}).

\section{Estimates of $\BBGAMAXS$ in the SM and the Parameters $(p_F,m_q)$ }
It is theoretically preferable to calculate the
branching ratio for the inclusive decay $\BGAMAXS$ in terms of the
semileptonic decay branching ratio
\begin{equation}
\label{brdef}
\BR ( B \ra  X_{s} \g) = [\frac{\Gamma(B \ra
\gamma + X_{s})}{\Gamma_{sl}}]^{th}
\, \BR (B \to X\ell \nu_\ell)  \qquad ,
\end{equation}
where, in the approximation of including the leading-order QCD correction,
 $\G_{sl}$ is given by the expression
\begin{equation}
\label{widthsl}
\G_{sl} = \frac{G_F^2 \, |V_{cb}|^2}{192 \pi^3} \, m_b^5 \,
g(m_c/m_b) \, ( 1-2/3 \frac{\alpha_s}{\pi} f(m_c/m_b)) \quad .
\end{equation}
The phase space function $g(z)$ is defined as
\begin{equation}
 g(z)=1-8z^2 +8z^6-z^8-24z^4\ln (z)  \quad ,
    \label{e7}
\end{equation}
and the function $f(z)$ can be seen, for example, in ref.
\cite{Alipiet}. An approximate analytic form for $f(z)$ has been
obtained in \cite{CCM}:
\begin{equation}
f(z) = (\pi^2-\frac{31}{4})(1-z)^2 + \frac{3}{2} \quad .
\end{equation}
To get the branching ratio in eq. (\ref{brdef}) we can proceed
in two ways. One can either take the partonic (purely perturbative)
expressions for both $\Gamma(B \to X_s \gamma)$ and
$\Gamma_{sl}$ or one can first implement the wave function effects and then
 integrate the spectra.
In the latter case, the dominant wave function effects
in the quantity $\Gamma_{sl}$
are included if one identifies $m_b$ in eq. (\ref{widthsl})
with the
effective value $W_{eff}$ of the $b$ quark mass, which is the value
of the floating $b$-quark mass averaged over the Gaussian distribution:
\be
W^5_{eff} = \int \, dp \, p^2 \, \phi(p) \, W^5(p) \quad ,
\ee
where $W(p)$ is given in eq. (\ref{lett14}).
We remark that these procedures yield an almost
identical branching ratio (within $1\%$), which shows that the
Gaussian-distributed Fermi motion model \cite{Alipiet,ACCMM}
 is in agreement with the result that
 power corrections to the inclusive decay widths
$\Gamma(\BGAMAXS )$ and $\Gamma (B \to X \ell \nu_\ell)$, calculated in
the context of the heavy quark effective theory \cite{HQETpower},
 cancel out in their ratio.

\indent
 We now estimate
  $\BBGAMAXS$ in the standard model and theoretical uncertainties on this
quantity.
 The parameters that we have used in estimating
the inclusive rates for $\BBGAMAXS$  are
 summarized in table \ref{tabparam}.
\begin{table}
\begin{center}
\begin{tabular}{| c | c | }
\hline
 Parameter & Range\\
\hline \hline
$\overline{m_t}$ (GeV) & $170 \pm 11$\\
$\mu$ (GeV) & $5.0^{+5.0}_{-2.5}$ \\
$\Lambda_5$ (GeV) & $0.195 ^{+0.065}_{-0.05}$\\
${\cal B}(B \to X \ell \nu_\ell)$ & $(10.4 \pm 0.4)\%$\\
$m_c/m_b$ & $0.29 \pm 0.02$\\
$m_W$ (GeV) & $80.33$ \\
$\alpha_{\footnotesize{\mbox{QED}}}^{-1}$ & $ 130.0$\\
\hline
\end{tabular}
\end{center}
\caption{Values of the parameters used in estimating the branching
ratio $\BBGAMAXS$ in the standard model. The range of $\Lambda_5$
is taken from the present world average (corresponding to $\alpha_s(m_Z^2)
=0.117 \pm 0.005$, using the two-loop $\beta$-function
\protect\cite{PDG94}) and the semileptonic branching ratio from
\protect\cite{Gibbons}.} \label{tabparam}
\end{table}
  The largest theoretical
 uncertainty stems from the scale dependence of the Wilson coefficients
 as was already stressed in \cite{AGM93}. The extent of this variation
 is somewhat correlated with the value of the QCD scale parameter
 $\Lambda_5$ and the top quark mass. As given explicitly in the
 preceding section, the decay rate for $\BGAMAXS$
 depends on seven of the eight Wilson coefficients given earlier, once
 one takes into account the bremsstrahlung corrections and is not
 factored in terms of a single (effective) coefficient, that one encounters
 for the two-body decays $b \to s + \gamma$ \cite{Buras94}.
 To get some insight in the errors we enumerate here
 the values of the two dominant effective coefficients, $C_7^{eff}$
 and $C_8^{eff}$, as one varies $\mu$, $\Lambda_5$ and $m_t$
 in the range given in table 2:
 \begin{eqnarray}
 C_7^{eff} &=& -0.306 \pm 0.050 \nonumber\\
 C_8^{eff} &=& -0.146 \pm 0.020
\label{c78eff}
\end{eqnarray}
The present theoretical uncertainties on these coefficients represent
the dominant contribution to the theoretical error on
$\BBGAMAXS$, contributing about $\pm 35 \%$. The second source of
 scale-dependence is due to the appearance of the running
 quark masses in the operators $O_7$ and $O_8$. As discussed
 in section 3, this brings into fore the extra (scale-dependent)
 multiplicative  factor
$R=(m_b(\mu)/m_{b,pole})^2$ for the branching ratio $\BBGAMAXS$.
 The next largest
error arises from the parameters which are extrinsic to the decay
$\BGAMAXS$ but have crept in due to our procedure of normalizing the
branching ratio $\BBGAMAXS$ in terms of the semileptonic branching
ratio. To estimate this extrinsic error, we note that the
                 ratio $m_c/m_b$ can be written as
$m_c/m_b = 1 - (m_b - m_c)/m_b$. The mass difference $m_b - m_c$ is
known to a very high accuracy through the $1/{m_Q}$ expansion
\cite{HQET2} or from the semileptonic $b \to c$ spectrum;
 to a rather high accuracy its value is
$m_b-m_c = 3. 40$ GeV \cite{Shifmanetal}.
The b-quark mass $m_b$ is, however, not known so precisely.
Using for the $b$ quark pole mass $m_{b,pole}=4.8 \pm 0.15$ GeV, one gets
$m_c/m_b=0.29 \pm 0.02$, which is consistent with
the determination of the same from the lepton energy spectrum
$m_c/m_b = 0.316 \pm 0.013$
\cite{Rueckl} but less precise. This introduces considerable uncertainty
in the theoretical estimates of the branching ratio $\BBGAMAXS$,
which can be judged from the numerical value $g(z=0.29 \pm 0.02)=
0.542 \pm 0.045$, leading to $\pm 8.1\%$ error on the branching
ratio $\BBGAMAXS$.
               We recall here that $f(z)$ is a slowly varying
function of $z$, and for the quark mass ratio relevant for the decay
                   $b \to c + ~\ell ~\nu_\ell$,
namely $z=0.29 \pm 0.02$, it has the value $f(z)=2.57 \mp 0.06$.
 Taking into account the
experimental error of $\pm 4.1 \%$ on $\BR(B \to X \ell \nu_\ell)$
\cite{Gibbons}, and adding these errors linearly, one gets an error of
$\pm 12 \%$ on $\BBGAMAXS$ from the semileptonic decays. The procedure
that we have adopted in estimating the theoretical uncertainties on the
inclusive branching ratio $\BBGAMAXS$
  is as follows: We propagate
the errors due to the scale $\mu$, the QCD scale parameter, $\Lambda_{5}$,
and the top quark mass in our calculations. As remarked already, this
 constitutes the largest error.
The extrinsic errors (from $m_c/m_b$ and
the semileptonic branching ratio), being obviously uncorrelated, are then
added in quadrature in quoting the branching ratio.

\indent
We now proceed to discuss our results.
Assuming $|V_{ts}|/|V_{cb}=1$ \cite{PDG94},
 we plot in Fig. 1 the branching
ratio $\BR(B \to X_s \g)$ as a function of the top quark mass
$m_t$.
\begin{figure}[htb]
\vspace{0.10in}
\centerline{
\epsfysize=3in
\rotate[r]{
\epsffile{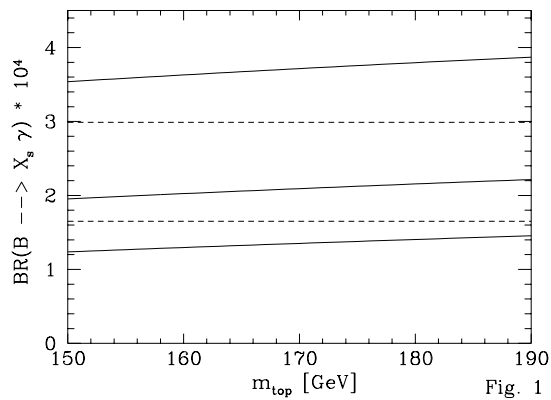}
}
}
\vspace{0.08in}
\caption[]{
$\BR(B \to X_s \g)$ as a function of $m_t$. The three solid lines
correspond to the variation of the parameters
 $(\mu,\Lambda_5)$ as descibed in the text.
The experimental $(\pm 1\sigma)$-bounds
 from CLEO \cite{CLEOrare2} are shown by the dashed lines.
\label{fig:1}}
\end{figure}
For all three solid curves, representing the SM-branching ratio,
 we have used $z=0.29$. The top
solid curve is drawn for $\mu=2.5$ GeV and $\Lambda_5 = 0.260$ GeV.
The bottom solid curve is for $\mu=10$ GeV and $\Lambda_5 = 0.145$
GeV,
 and the  middle solid curve corresponds to the central values of
the input parameters in table 2.
 Using $\overline{m}_t=(170 \pm 11)$ GeV, and adding the extrinsic error,
 we get
\be
\BBGAMAXS=(2.55 \pm 1.28) \times 10^{-4} \,,
\ee
 to be compared with the
CLEO measurement $\BBGAMAXS = (2.32 \pm 0.67) \times 10^{-4}$.
The $(\pm 1\sigma)$-upper and -lower bound from the CLEO measurement
are shown in Fig. 1 by dashed lines. We see that the agreement between SM
and experiment is good. The theoretical errors estimated by us are, however,
larger, than for example in \cite{Buras94}, for reasons that we have
explained above.

In Fig. 2 we show the branching ratio $\BBGAMAXS$ as a function
of the CKM matrix element ratio squared $|V_{ts}|/|V_{cb}|^2$, varying
$m_t$, $\mu$ and $\Lambda_5$ in the range specified in table 2.
\begin{figure}[htb]
\vspace{0.10in}
\centerline{
\epsfysize=3in
\rotate[r]{
\epsffile{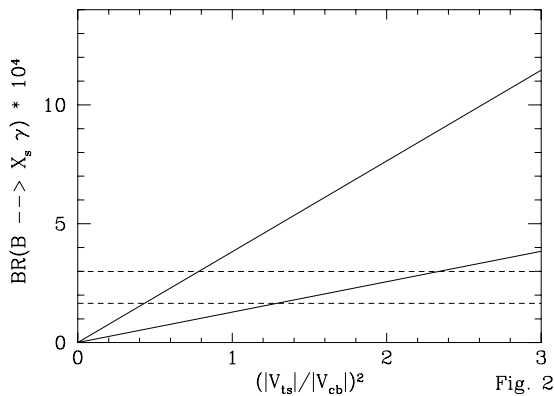}
}
}
\vspace{0.08in}
\caption[]{
$\BR(B \to X_s \g)$ as a function of $(|V_{ts}|/|V_{cb}|)^2$.
The solid lines
correspond to the variation of the parametrs
 $(\mu,\Lambda_5,m_t)$
in the limits specified in table 2.
The experimental $(\pm 1\sigma)$-bounds
 from CLEO \cite{CLEOrare2} are shown by the dashed lines.
\label{fig:2}}
\end{figure}
Using the $(\pm 1 \sigma)$-experimental bounds on the branching ratio (dashed
lines) we infer
\be
 |V_{ts}|/|V_{cb}|=1.10 \pm 0.43 \,,
\ee
which is consistent with the indirect constraints from the CKM
unitarity \cite{PDG94}.

 Now we discuss the photon energy spectrum and fit of the
parameters  $p_F$ and $m_q$
by  using the CLEO data which has been corrected for the detector effects.
The  photon yield from the decay $\BGAMAXS$ is given in table 3 in photon
energy bins having a width of 250 MeV starting with $E_\gamma =1.95$ GeV.
We note that these entries  are based on the weighted average of the two
different
methods (event shape and B reconstruction) used by the CLEO collaboration
in the analysis of their data
 \cite{CLEOprivate}. The photon energy has been measured in the laboratory
frame (i.e. in the rest frame of $\Upsilon (4S)$) and
the numbers in table 3 are presented  in this frame.
\begin{table}
\begin{center}
\begin{tabular}{| c | c | }
\hline
 $E_\gamma$-interval  & number of events \\
\hline \hline
1.95 - 2.20 GeV & $229 \pm 256$\\
2.20 - 2.45 GeV & $484 \pm 163$\\
2.45 - 2.70 GeV & $381 \pm 105$\\
2.70 - 2.95 GeV & $12  \pm  59$\\
\hline
\end{tabular}
\end{center}
\caption{Photon yield in the laboratory frame from the decay $\BGAMAXS$,
obtained from the
measurement of the photon energy spectrum by the CLEO collaboration
\protect\cite{CLEOrare2} based on a sample of 2.152 million
$B \bar{B}$ events.
The data shown have been corrected due to the
detector acceptance \protect\cite{CLEOprivate}.}
\label{tabgambin}
\end{table}
This implies that the $B$ mesons from the $\Upsilon(4S)$
decay have a momentum of
$\approx
350$ MeV, and in doing the analysis we have boosted the theoretical
rest-frame spectra accordingly.
As the theoretical uncertainties are mainly in the normalization
of the spectrum, we normalized both the theoretical predictions
(parametrized in terms of $p_F$ and $m_q$) and the experimental
data to unit area in the interval between 1.95 GeV and 2.95 GeV.
We then performed a $\chi^2$ analysis. The experimental
errors are still large and the fits result in relatively small $\chi^2$
values; the minimum, $\chi^2_{min}=0.038$, is obtained
for $p_F=450$ MeV and $m_q=0$, which corresponds to the $b$-quark pole mass
$m_{b,pole}=4.77$ GeV, in good agreement with theoretical estimates of
the same.
 A  contour plot
in the $(m_q,p_F)$ plane, obtained by varying the $\chi^2$ by one unit
from $\chi^2_{min}$,
 which corresponds to
 $\approx 39 \%$ confidence level \cite{PDG94}, is shown in Fig. 3.
\begin{figure}[htb]
\vspace{0.10in}
\centerline{
\epsfysize=3in
\rotate[r]{
\epsffile{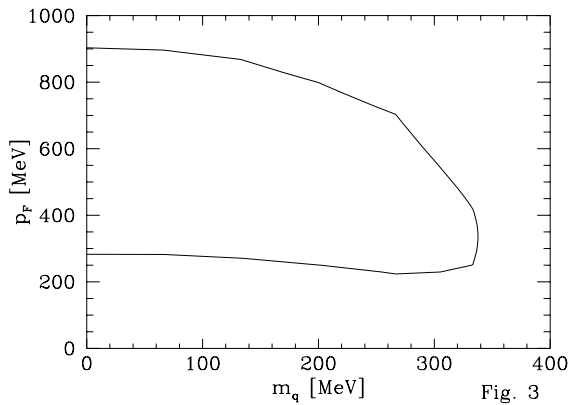}
}
}
\vspace{0.08in}
\caption[]{
 Contourplot in the $(m_q,p_F)$ parameter space obtained from $\chi^2=
\chi^2_{min} +1$.The  minimum $\chi^2$ is for the values (0 MeV, 450 MeV).
\label{fig:3}}
\end{figure}
As expected, due to the large errors in data the parameter space
cannot be restricted very much at present. Nevertheless,
the results for $p_F$ are compatible with  the value $p_F=270 \pm 40$ MeV
obtained from the $B$-semileptonic decay analysis \cite{CLEOrare2}.
In Fig. 4 we have plotted the photon energy spectrum normalized
to unit area in the interval between 1.95 GeV and 2.95 GeV for
the parameters which correspond to the minimum $\chi^2$ (solid curve)
and for another set of parameters that lies near the
$\chi^2$-boundary in the contour plot
(dashed curve). Data from CLEO \cite{CLEOrare2} are also
shown. \begin{figure}[htb]
\vspace{0.10in}
\centerline{
\epsfysize=3in
\rotate[r]{
\epsffile{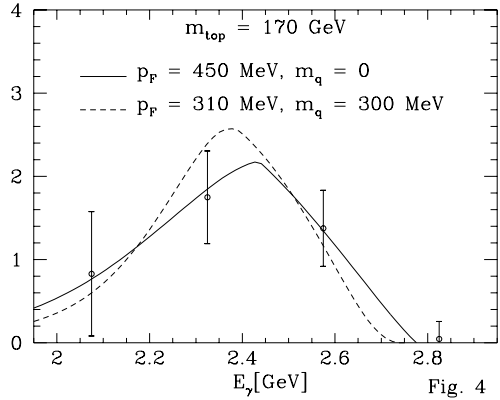}
}
}
\vspace{0.08in}
\caption[]{Comparison of the normalized photon energy distribution
using the corrected CLEO data \protect\cite{CLEOrare2} and our theoretical
distributions, both  normalized to unit area in
the photon energy interval between 1.95 GeV and 2.95 GeV. The solid
 curve corresponds to the values with the minimum $\chi^2$,
 $(m_q,p_F)$=(0,450 MeV), and the dashed curve to the values
 $(m_q,p_F)$=(300 MeV, 310 MeV).
\label{fig:4}}
\end{figure}

In Fig. 5 we show the comparison of the differential
branching ratio from CLEO with our calculations. The
theoretical  curves correspond to the central
values
of the input parameters in table 2. The agreement between experiment and
SM is good.
\begin{figure}[htb]
\vspace{0.10in}
\centerline{
\epsfysize=3in
\rotate[r]{
\epsffile{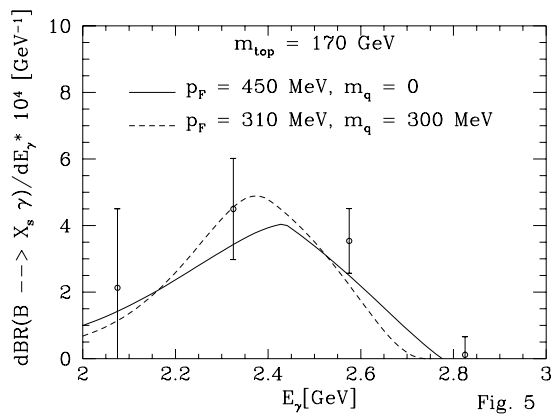}
}
}
\vspace{0.08in}
\caption[]{The same as in
 Fig 4, but with absolute (i.e not normalized) differential branching ratio.
\label{fig:5}}
\end{figure}

  In summary, we have presented an improved theoretical calculation of the
branching ratio $\BBGAMAXS$ and the photon energy spectrum
in $\BGAMAXS$, using
 perturbative QCD and a Gaussian Fermi motion model. We estimate
$\BBGAMAXS = (2.55 \pm 1.28) \times 10^{-4}$ in the SM, in agreement with
the corresponding branching ratio $\BBGAMAXS =(2.32 \pm 0.67) \times 10^{-4}$
reported by the CLEO collaboration. Our model calculations provide a good
account of
the measured photon energy distribution, and the best-fit parameters
correspond to $p_F=450$ MeV and $m_{b,pole}=4.77$ GeV. The errors on these
parameters are still large but within errors both $p_F$ and $m_{b,pole}$
determined from the radiative and semileptonic $B$ decays are compatible
with each other.
Precise comparison  requires improved
measurements and theory, which we hope are forthcoming.

{\bf Acknowledgements}
  We are very grateful to the members of the CLEO collaboration, in
particular Tomasz Skwarnicki and Ed Thorndike, for providing
table 3 and
for numerous helpful discussions. We acknowledge helpful correspondence
and discussions with Matthias Neubert and Vladimir Braun on  quark
masses. Discussions with Guido Martinelli, Thomas Mannel,
 Giulia Ricciardi and Daniel Wyler are also thankfully acknowledged.
We also thank Frank Cuypers for providing us with a program to
draw the contour plot and for useful general discussions on statistics.
One of us (C.G.) would like to thank the DESY theory group for its
hospitality.



\begin{thebibliography}{99}

\bibitem{CLEOrare2}
M.S. Alam et al. (CLEO Collaboration), Phys. Rev. Lett. {\bf 74} (1995) 2885.
\bibitem{CLEOrare1}
 R. Ammar et al. (CLEO Collaboration), Phys. Rev. Lett. {\bf 71} (1993) 674.
\bibitem{ag1}  A. Ali and C. Greub,
              Z. Phys. {\bf C49} (1991) 431;
              Phys. Lett. {\bf B259} (1991) 182.
\bibitem{ag2}  A. Ali and C. Greub ,
               Phys. Lett. {\bf B287} (1992) 191.
\bibitem{ag3}  A. Ali and C. Greub,
              Z. Phys. {\bf C60} (1993) 433.
\bibitem{Buras94}
       A.J. Buras, M. Misiak, M. M\"unz, and S. Pokorski,
       Nucl. Phys. {\bf B424} (1994) 374.
\bibitem{Ciuchini94}
       M. Ciuchini et al., Phys. Lett. {\bf B334} (1994) 137.
\bibitem{Ciuchini}
     M. Ciuchini et al.,
     Phys. Lett. {\bf B316} (1993) 127; Nucl. Phys. {\bf B415} (1994) 403;\\
     G. Cella et al., Phys. Lett. {\bf B325} (1994) 227. \\
     M. Misiak, Nucl. Phys. {\bf B393} (1993) 23;
     Erratum ibid. {\bf B439} (1995) 461.
\bibitem{Grinstein90}
     B. Grinstein, R. Springer, and M.B. Wise, Phys. Lett. {\bf 202}
                  (1988) 138; Nucl Phys. {\bf B339} (1990) 269.
\bibitem{Alipiet} A. Ali and E. Pietarinen,
                  Nucl. Phys. {\bf B154} (1979) 519.
\bibitem{ACCMM} G. Altarelli et al.,
                  Nucl. Phys. {\bf B208} (1982) 365.
\bibitem{InamiLim}
        T. Inami and C.S. Lim,
        Prog. Theor. Phys. {\bf 65} (1981) 297.
\bibitem{AGM94}
A. Ali, G. Giudice, and T. Mannel, CERN-TH.7346/94 and Erratum
(to appear in Z. Phys. C).

\bibitem{AGlong}
        A. Ali and C. Greub, DESY Report (in preparation).

\bibitem{CCM}
G. Corbo, Nucl. Phys. {\bf B212} (1983) 99; N. Cabibbo, G. Corbo,
and L. Maiani, {it ibid.} {\bf B155} (1979) 93.

\bibitem{JK}
M. Jezabek and J.H. K\"uhn, Nucl. Phys. {\bf B320} (1989) 20.

\bibitem{Sudakov}
V. Sudakov, Zh. Eksp. Teor. Fiz. {\bf 30 }(1956) 87
            [ Sov. Phys. JETP {\bf 3} (1956) 65 ] .

\bibitem{Neubert94}
M. Neubert, Phys. Rev. {\bf D49} (1994) 4623.

\bibitem{shifmangamma}
R.D. Dikeman, M. Shifman, and R.G. Uraltsev, Preprint TPI-MINN-95/9-T,
UMN-TH-1339-95, UND-HEP-95-BIG05 (hep-ph/9505397).

\bibitem{HQETpower}
J. Chai, H. Georgi, and B. Grinstein, Phys. Lett. {\bf B247} (1990) 399;\\
I. Bigi, N. Uraltsev, and A. Vainshtein, Phys. Lett. {\bf B293} (1992) 430;
{\bf B297} (1993) 477 (E);\\
A. Falk, M. Luke, and M. Savage, Phys. Rev. {\bf D49} (1994) 3367.

\bibitem{AGM93}
A. Ali, C. Greub and T. Mannel, DESY Report 93-016 (1993), and in
{\it $B$-Physics Working Group Report,
     ECFA Workshop on a European $B$-Meson Factory}, eds.: R. Aleksan
     and A. Ali, ECFA-Report 93/151, DESY 90-053 (1993).

\bibitem{PDG94}
  L. Montanet et al. (Particle Data Group), Phys. Rev. {\bf D50} (1994)
1173.

\bibitem{Gibbons}
L. Gibbons (CLEO Collaboration), in Proceedings of the XXX Rencontres
de Moriond, Les Arcs, March 1994.

\bibitem{HQET2}
 I. Bigi et al., Phys. Rev. Lett. {\bf 71} (1993) 496.

\bibitem{Shifmanetal}
M. Shifman, N.G. Uraltsev, and A. Vainshtein, Phys. Rev. {\bf D51}
(1995) 2217;\\
M.B. Voloshin, TPI-MINN-94/38-T (hep-ph/9411296).

\bibitem{Rueckl}
R. R\"uckl, MPI-Ph/36/89.

\bibitem{CLEOprivate}
 E.H. Thorndike and T. Skwarnicki (private communication).
 \end{thebibliography}
\end{document}